# From Bitcoin to Solana – Innovating Blockchain towards Enterprise Applications

Xiangyu Li, Xinyu Wang, Tingli Kong, Junhao Zheng and Min Luo

Georgia Institute of Technology, Atlanta, GA 30332, USA
`mluo60@gatech.edu`

**Abstract.** This survey presents a comprehensive study of recent advances in blockchain technologies，focusing on how issues that affecting the enterprise adoption were progressively addressed from the original Bitcoin system to Ethereum, to Solana etc. Key issues preventing the wide adoption are scalability and performance, while recent advances in Solana has clearly demonstrated that it is possible to significantly improve on those issues by innovating on data structure, processes and algorithms by consolidating various time-consuming algorithms and security enforcements, and differentiate and balance users and their responsibilities and rights, while maintaining the required security and integrity that blockchain systems inherently offer.

**Keywords:** Blockchain, Distributed Ledger, Consensus, Proof of Work, Proof of History Scalability, Performance, Security,

## 1 Introduction

### 1.1 Rise of Blockchain Technology

The blockchain is a purely distributed peer-to-peer system of ledgers that utilizes some well-articulated software constructs of algorithms, collaboratively peers to record and negotiate the informational content of ordered and connected blocks of transaction data together with cryptographic and security enrichments to achieve integrity. It was first introduced by Bitcoin in 2009 and has been becoming a mainstream technology. It has been used in various industries, such as financial, healthcare, supply chain, logistics, and many others. Such distributed ledgers are designed to provide a permanent, tamper-proof record of business transactions, as they can be utilized to improve collaboration, enable provenance, speed up transaction settlements or enable transparency.

Blockchain can also be viewed as a decentralized database running on a peer-to-peer network, where each node/computer (or some selected group) maintains a copy of the current ledger. It offers data security and reliability as the data cannot be easily modified while the redundant copies make data loss unlikely.

Blockchain innovated in how digital information is stored, verified and exchanged, and was inherently designed and developed to create secure, reliable and transparent business processes for enterprises. One of the surveys reveals that the global block-chain market size is expected to grow from USD 3.0 billion in 2020 to USD 39.7 billion in 2025, at a CAGR of 67.3% during 2020-2025 [1]. As organizations have started to explore and experiment with blockchain's potential by developing blockchain applications, the proper choice of a "good" blockchain platform becomes vital. As they become increasingly more popular, enterprises need better information to make right judgement calls to decide not only when to jump into the tech wagon, but more importantly how they can take advantage of the new technology while avoiding potential pitfalls.

Blockchain and smart contracts make it possible for multiple parties to share business logic and collaboratively conduct business processes/operations automatically. Properly utilized, it can reduce IT costs, expand B2B and B2C networks, enable new products and service that could bring in revenue and profits. Moreover, blockchain's business value is expected to increase as enterprise implementations proliferate and are further extended and refined.

## 1.2 Issues facing Enterprise Adoption of Blockchain Technology

This paper will not cover whether blockchain technology fits enterprises from the business perspective, although that should be the first question to ask. We will focus on non-functional requirements that describes the system's operation capabilities and constraints that enhance its business functionality. The non-functional requirements needed for application will of course depend on the business context and the outcomes to be achieved, particularly as there are so many that can be applied. In this paper, we will only elaborate a few most critical ones.

**Performance.** All enterprise systems should be designed and built with an acceptable standard of performance as a minimum, while taking into accounts problems such as scalability, latency, load and resource utilization. Many factors could negatively impact performance, including high numbers of API calls, poor caching, and high-load third-party services. It's critical to ensure the end-user experience or integration of multi-systems across the entire eco-chain is not affected by any such issues.

Prevailing business transaction systems have been capable of processing thousands (Visa, for example) or millions of transactions (online market place such as Amazon or Alibaba) per second without any failure, most of the current blockchain platforms depicted a remarkable slowdown, making them unviable for large-scale or performance-sensitive applications. For example, Bitcoin can only process roughly 3 to 7 transactions per second, with Ethereum about 15 to 20 transactions.

Such poor performance and cumbersome operations are mainly due to the complexity with encrypted and distributed nature in blockchains. Although it is not at all suitable for high-frequency transactions, ways to improve its transaction performance, including throughput and latency, is always a hot topic. Compared to "traditional" payment systems such as cash or debit cards, it could take hours or even days to process some transactions. When more users join the network, its performance will be further degraded due to the existence of consensus latency from nodes with low processing power. As a result, the transactions cost is higher than usual, further limiting more users onto the network.

**Scalability.** Scalability is the second big issue that needs to be addressed, as this is one of the core reasons why organizations still hesitate to adopt blockchains.

The system must be able to accommodate ever-increasing volumes (number of users/devices/integrated applications, data and throughput) over time, and is able to scale up and down quickly as the number of users change drastically, as needed.

**Security and Integrity.** Requirements such as confidentiality, authentication and integrity ensure that valuable (private and confidential) information is protected. Blockchain benefits primarily derive from the trust it fosters, its built-in privacy, security and data integrity and its transparency, as it incorporates a flow of data from complex mathematical operations that cannot be changed once created without being detected, and every transaction is encoded and connected, and therefore it is significantly more reliable than traditional journal methods. This unchangeable and incorruptible characteristic inherently make blockchains safer and better protected against tampering and hacking of information.

Various software engineering tactics can be employed to safeguard valuable/transactional data at many integration points. System architects need to understand legal and compliance requirements and communicate these clearly to the development team, so that the necessary levels of security can be established and enforced jointly.

With blockchains, an external audit can be provided from the distributed ledger. This will inherently enhance privacy and avoid corruption, and help confirm the legitimacy of transactions and offer indisputable proof of transactions.

**Availability/Reliability/Resilience.** The system must be available for use, and the downtime must be reduced to an acceptable level under any circumstances. For example, mechanisms to avoid single points of failures, and adequate timeouts could be used to enhance system availability and reliability.

**Feasibility.** Feasibility considers issues such as technology maturity, time-to-market, total cost of ownership, technical knowledge, and migration requirements. Commercial-off-the-shelf (COTS) solutions, managed services and cloud-native functions where appropriate, and close collaboration with development partners with suitable architecture and solution components and services will definitely help address those issues. .

This paper surveyed several important blockchain platforms covering the years of evolution from the original Bitcoin system to the more advanced recent offerings. Hopefully, with the information we collected and analyzed, it could help enterprises to make better decisions, while also directs new players where to innovate in order to make blockchain well fit into most enterprises business needs. We will review the chosen frameworks, especially their data structures, processes and algorithms involved in creating a new transaction record (block), and how conflicts or disputes could be resolved in Section 2. We will also raise concerns on several key issues related to the afore-mentioned NFRs, especially the performance and scalability. Section 3 then proceeds to analyze the selected platforms and discuss, from the evolu-tionary nature of blockchain technology since its inception, how critical issues such as performance, and scalability etc. were addressed, especially the most recent advances from Solana, where 2 -4 order of magnitude of improvements has been proved possible. Section 4 will first present a quick summary view on how enterprises could leverage the information collected and analyzed in the maker to choose "better-fit" platforms, then points to some remaining issues that should be further evolved or even revolutionized to truly meet some fundamental NFR requirements for enterprise adoption. Some alternative approaches to achieve the secure and immutable nature of the distributed ledger is also included. Section 5 conclude the paper with a few quick remarks.

## 2    Main Frameworks and Consensus Algorithms

This section will give a general description of blockchain architecture, in terms of how blocks are structured and organized into a chain. Issues related to consensus, performance, and scalability will be explored respectively. The reasons for choosing these five platforms as examples, including Bitcoin, Ethereum, Hyperledger Fabric, EOS and Solana, will be explained at the end of this section.

### 2.1    General Description of Blockchain and Its Main Data Structure

Blockchain is a chain linking or "chaining" different blocks, while a block is the foundation and formed by recording and calculating all the transactions in a Merkle tree and adding the previous block header hash value(s) into the current header, as in Fig. 1.

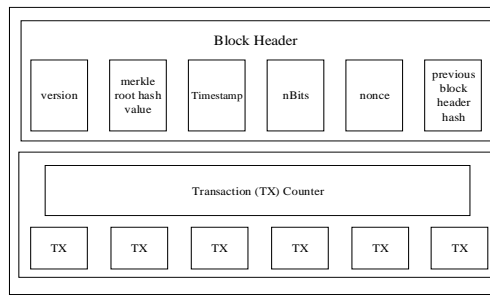

**Fig. 1.** Block Structure

The hash value of the previous block will be included in the current block hash. Fig. 2 shows how one block is connected to the other. Note that the first block only has the hash value from its own transactions [2].

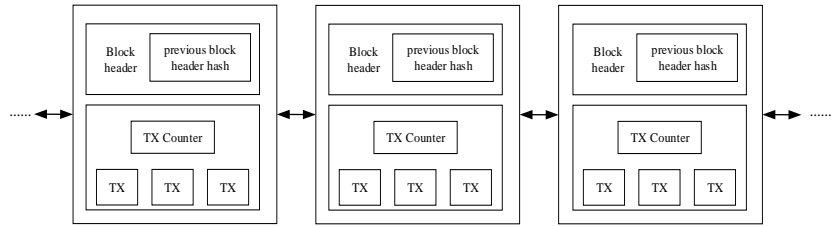

**Fig. 2.** Chain of block bodies

The main body of each block is structured as a Merkle Tree in Fig. 3, where every transaction is first hashed individually and its hash value is then hashed with another hash value.

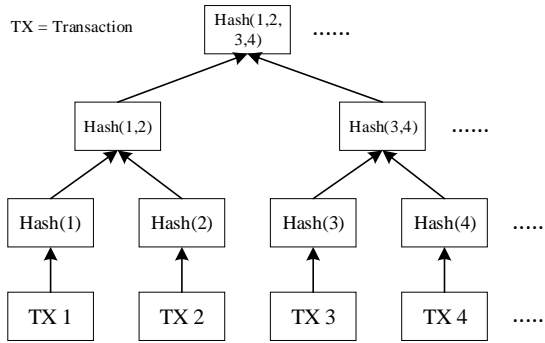

**Fig. 3.** Merkle Tree inside block body

**Smart contract**. In blockchain network, the smart contract is "a secure and unstoppable computer program representing an agreement that is automatically executable and enforceable" [3]. It materializes rules, definitions and expectations in the forms of code and data, so that all nodes will act accordingly.

The smart contract should be executed by all nodes in the network and the same results can be obtained. For instance, during transaction process, the smart contract inside nodes performs calculations, stores information, publicizes state and does transferring. Every node has no choice but to obey these rules and get same results. Under such circumstances, a referee or a third party is not needed. Thus, the crisis of trust inside the network is largely decreased.

When implementing smart contracts at the enterprise level, a co-evolution of both contract and technology is highly required. Enterprise smart contracts can provide a series of service by modularizing data, contract participants and external dependencies [4]. These services can perfectly satisfy the requirements of privacy, scalability and internal administration. More values can be realized in smart contracts with shared and cross-organizational environments that could be enabled by blockchain technology.

**Consensus issues.** The consensus algorithm is a mechanism that ensure all distributed untrustworthy nodes keep the same ledger by making recorded transactions immutable and maintain consistent states. By impartially verifying and validating transactions, the nodes will be rewarded according to their efforts in this process. Two core proof-based algorithms, PoW and PoS [5], bring in some basic issues to be addressed by later versions of consensus algorithms.

*Proof of Work (PoW).* Proof of Work encourages nodes or users in the network to devote their computational power for transaction process by rewarding them for their efforts [6]. If one node initiates a block of transactions, this block will have to be checked with computations by all other nodes, the so called mining process while the participating nodes are miners. Miners will contain a nonce when working out a hash value. This value will eventually be tried out by adjusting nonce and thus a block is validated.

Such consensus mechanism will cause huge waste of computing resource. The whole network of miners will spend their best effort in working out only one hash value. Except the miner who first works it out and gets rewarded, other miners only just wastes their computing power. The efficiency of the mechanism is also low. The time for the mining process would be around 10 minutes with only one output, etoo low for the real-world business transactions.

*Proof of Stake (PoS).* Proof of Stake is based on the amount of balance each miner possesses. As many miners may find validated blocks easier with comparatively more computing ability, PoS is designed by rewarding miners with interests based on the amount they own [7]. Their possessions are the "stakes", and it is the stake that decide who will mine the following blocks. There is no competition among miners, and therefore computational waste is reduced to some extent.

However, this mechanism is unreliable. As interests will be rewarded, some miners will large amount of stake might be unwilling to contribute their computational ability and rely only on stakes. This is negative trend that will give rise to lower mobility of transactions.

**Performance issues.** From a technical point of view, the typical blockchain network, such as Bitcoin and Ethereum, requires consensus from all nodes in the whole network. Even if a node completes its validation process, it has to wait for consents from other nodes. For Bitcoin, the throughput rate is 7 transactions per second (TPS) and the confirmation time is 60 minutes. Ethereum blockchain has a better performance with dozens TPS. Such throughput cannot satisfy large-scale enterprise applications. This issue will be even more acute when more users/nodes join the network.

**Scalability issues.** The processing power of individual nodes largely determines the scalability of the blockchain system. For instance, when it comes to Bitcoin and Ethereum, each core node in the network that participates in maintenance should maintain a complete storage and be processed.

Many other issues also impact the maturity and the adoption of the blockchain technology, including security and privacy, interoperability, availability and resilience, etc. However as indicated earlier, this paper will focus on the above more critical non-functional related requirements.

## 2.2 Types of Blockchain Platforms

As more businesses look for adopting blockchain technology, various blockchain platforms have been developed that can be categorized by how open or closed they allow participants contribute to business transactions or verify the accuracy of each block added to the blockchain and the distributed ledger.

All types of blockchains can be characterized as permissionless, permissioned, or both. Permissionless blockchains allow any user to pseudo-anonymously join the blockchain network with full rights, while permissioned blockchains restrict access and also rights to certain nodes. Permissionless blockchains tend to be more secure and reliable than permissioned blockchains, while permissioned blockchains tend to be more efficient, as access to the network is restricted with fewer nodes on the blockchain system.

**Table 1.** Summarized key features and pros/cons of the four types.

|  | *Public* | *Private* | *Consortium* | *Hybrid* |
|---|---|---|---|---|
| *Perminssion* | Permissionless | Permissioned | Permissioned | Both |
| *Advantage* | Independece | Access Control | Access Control | Access Control |
|  | Transparency | Performance | Security | Performance |
|  | Trust | Scalability | Scalability | Scalability |
|  |  |  |  | Limited Independence and Transparency |
| *Diadvantage* | Performance: Long validation times | Security & Trust: More vulnerable to fraud and bad actors | Transparency | Transparency |
|  | Scalability | Auditability | Improved Security & Trust | Upgrading |
|  | Security |  |  |  |
| *Typical Use Cases* | Cryptocurrency | Supply chain | Banking | Medical |
|  | Doc validation | Asset Ownership | Research | Real Estate |
|  |  |  | Supply Chain |  |
| *Example Chains* | Bitcoin、Litecoin | Ripple：virtual currency exchange network | R3: financial services | BM Food Trust：whole food supply chain |
| | Ethereum | Hyperledger：General open-source blockchain applications | CargoSmart - Global Shipping Business Network Consortium，shipping industry | |

**Public (permissionless)** blockchain opens to ALL, while not requiring any permission to join. Its consensus process involves all nodes that makes data verification very tedious and time consuming, but it also make the system less vulnerable to hacking or control by a dominant actor. Cryptocurrency uses such chains.

**Private (permissioned,** managed) blockchain runs on a private network and could be controlled by a single organization, the central authority. It also has the same peer-to-peer architecture as public blockchain, but with significantly reduced scale and therefore better performance. But due to the nature of central/control node(s), its trust is weaker than public blockchains. Security could also be weaker because a small number of nodes could easily decide the consensus used to validate transactions, negating the original intention of the blockchains. Many early blockchain deployments use private blockchains.

**Hybrid** blockchain combines the features of public and private chains. Such a chain is controlled by a single organization, but with some oversight performed by the public blockchain. It can be used to partition some data and transactions behind a permission scheme while maintaining connections to the public chains. By not allowing the owner to modify transaction data, the security and data integrity risks of private blockchain can be alleviated with potentially better performance than public chains.

**Consortium** blockchain is similar to private blockchains. It is controlled by a group instead of a single entity, therefore less security susceptible than private chains.

### 2.3    Why We Choose the Six Platforms?

This paper is about innovating blockchain technology for enterprise adoption that could revolutionize how businesses can take advantages of the inherent secure information exchange and transaction integrity, and make the end-to-end integration of cross-border, organizations and business units seamless, driven and managed by agreed upon contracts that can be automatically executed with trust-worthy results. As the number and quality of blockchain platforms with enterprise-class development tools and architectures has reached a point where most companies can find a suitable platform and supportive community of developers and system integrators, it is still essential to understand their underlying technology stack and related algorithms, their relative merits, in order to find the best possible match for future business growth.

After analyzing almost every available blockchain platforms in the current market place, we selected six, 2 representatives, Bitcoin and Ethereum, for mostly public blockchain; and 2 Hyperledger Fabric and EOS, for private or alliance chains. Number 5 is R3 Corda, a non-traditional blockchain based distributed ledger. The most recent news regarding some very innovative mechanisms introduced in Solana boasted a 2+ order of magnitude improvements on TPS, and we believe that it is really the perfect Number 6 that not only promising, but more importantly evidence that basic blockchain structures and algorithms could still be significantly innovated to serve as the foundation of many enterprise applications. Fig. 4 shows the evolution timeline of the five platforms.

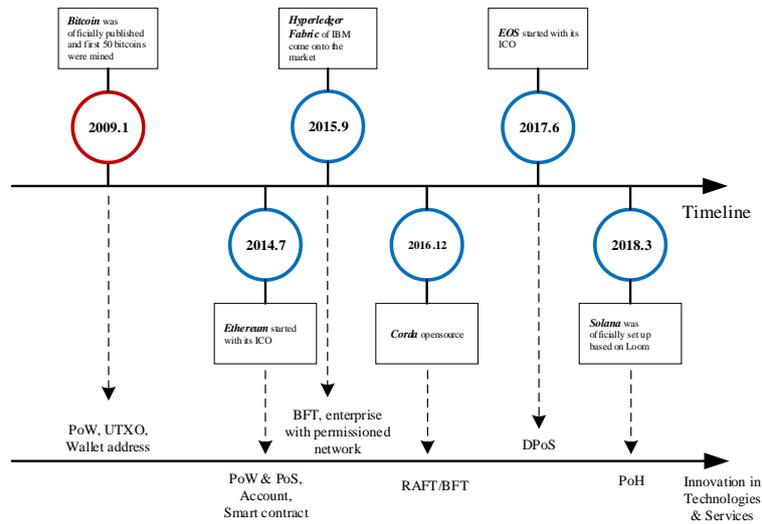

**Fig. 4.** Development of Blockchain platforms

Bitcoin and Ethereum are the top 2 well-known public blockchain platforms. Bitcoin is the first realization of blockchain and brings in consensus algorithms in a peer-to-peer system. Ethereum modify the traditional bitcoin structure by successfully implementing accounts and smart contracts. However, these public blockchains have their deficits in scalability and performance.

Under such circumstances, the private blockchain and consortium blockchain come into sight. Hyperledger Fabric and EOS are two enterprise blockchain platforms. In these permissioned networks, not all the nodes are equal peers – consensus verifying work is allocated among a small group of members. The consensus algorithms will also be less complicated than those in the public blockchain.

R3 Corda is a representative directed-acyclic-graph based distributed ledger with similar security and immutability as in basic blockchains (Bitcoin, Ethereum), but also better performance.

As mentioned, one of the latest blockchain platform with exciting news is Solana [8]. Solana considerably exceled in terms of its high transaction performance with improved consensuses. Hopefully, if Solana is adopted in enterprise blockchain instead of the existing platforms like Hyperledger Fabric and EOS, the company efficiency can be considerably increased.

Nevertheless, blockchain technology is still under development and man limitations are still need to be further exploited and ameliorated. We will discus these possibilities in the following sections.

## 3 Analysis of the Selected Main Frameworks

### 3.1 Bitcoin

Bitcoin was first designed to replace the use of "cash" in our real world, rather than for the enterprise-level system. Therefore, the concept of "wallet address" (or "wallet") is introduced. This is because Satoshi Nakamoto, the inventor, referenced the model of e-cash when designing the model. It is thus easy to understand that like cash can be put in many places, a Bitcoin user can have many wallets/addresses, with all amounts of balance inside each adding up to the total. These amounts of balance are called "unspent transaction outputs" (UTXO).

UTXO comes from inputs of transactions [9]. However, every input of UTXO is a separate entity which must be used up at a time. Multiple inputs can be inserted in to an address, while up to two outputs can be initiated each time, one for a targeted receiver, the other for getting back the remaining bitcoins. Also, a transfer can be initiated by changing UTXO's current address into the receiver's address. Only the sender with the private key can have access and transfer its UTXO to another address [10]. With UTXO, a transaction in Bitcoin network is just the change of balance's address.

Fig. 5 shows how different components of the Bitcoin system work in a transaction process. Before a transaction, network must verify from the previous record whether the sender has enough balance to send. After the validation from all miners (validators) in the whole network, only the first miner who figure out the output is rewarded.

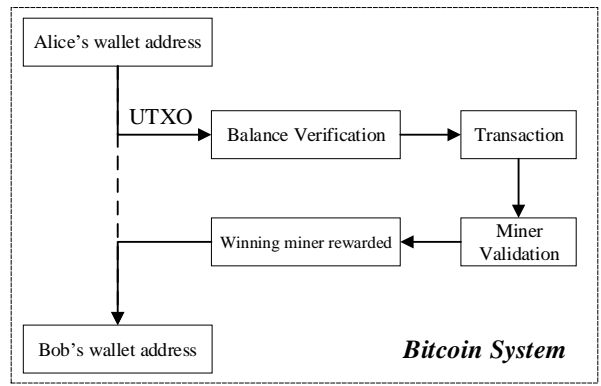

**Fig. 5.** Bitcoin system components

The above-mentioned transaction is recorded in transaction records. However, in Bitcoin, one deficit with transaction history is apparent: increased verification complexity. When one transaction is initiated, at most two members' previous transactions will be verified – one member is the receiver and the other is the sender. These two members will be traced back for previous transaction records for the amounts of balance in their wallets. As time passes by, the transaction records have been accumulating, and the verification process will be more and more complicated.

This case above is only a one-transaction scenario. If more nodes are added to Bitcoin, when doing transactions, transaction records from more nodes will be considered for calculating balance of a single node, and this will also increase verification complexity.

During transactions process, any changes to the transaction record is prohibited, otherwise the whole chain will be considered invalid. Theoretically, there are up to 5 illegal changes: data content change, Merkle-tree reference change, transaction substitution, Merkle-root change and block-header reference change [11].

The detection of changes is realized by checking changes in block header hash values. In Figure 6, we denote Transaction 1 to 4 as a part of the Merkle tree. If transaction 2 is changed or replaced, the value of R2 will also change, which will lead to the change in R12 and R34.

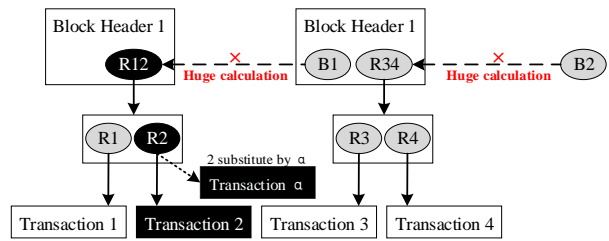

**Fig. 6.** Bitcoin illegal changes

To be more specific, once an element is changed, another element that points to it should also be changed, as the lower-level hash values will influence the higher-level ones.

### 3.2 Ethereum

Ethereum is a decentralized public ledger for verifying and recording transactions. The users of the network can create, publish, monetize, and use applications on the platform, and use its cryptocurrency "Ether" as payments. Two innovative concepts are introduced to Ethereum, smart contracts and account information.

EVM (Ethereum Virtual Machine) [12] creates an environment for smart contracts and makes it possible for anyone to create his own contracts and decentralized applications (DApps) [13]. This involves the definition of ownership rules, transaction methods and state transition functions. Smart contracts can further be expanded into business and enterprise level. The codes and data inside represent principles and rules that can be used to provide services according to different scenarios. Several factors can be taken into account, such as internal management, member conducts, privacy etc.

Besides, account [14] is also successfully implemented. There are two sorts of accounts in Ethereum: External Owned Accounts (EOA) [14] and Contract Accounts (CA) [14]. EOA is an e-cash account which encompass balance; while CA has both balance state and contract state. With accounts, the state information, such as balance of users, can be digitalized. This omits the needs to trace back transaction history for balance as in Bitcoin.

Ethereum transactions are validated data that an external account sends to another account [15]. There are three types of transactions: transactions that transfer value between two EOAs; transactions that send a message call to a contract; and transactions that deploy a contract. As all miners are rewarded in a transaction, Gas [16], which can be converted into Ether later, is introduced to restrict the usage of resources. Specifically, to take environment factors into account, such as bandwidth, computational complexity and storage space, Gas value is adjusted after current transaction for the next one.

Fig. 7 shows how the above smart contracts and account states are used in the transaction [17]. Before a new transaction starts, gas value is decided based on Ethereum network conditions. As Fig. 7 depicts, Alice first initiate the transaction and then broadcast the whole network. This transaction is added to a block and then miners begin validating. Every miner is reward for its effort with Gas, the amount of which depends on contributions in validating a block.

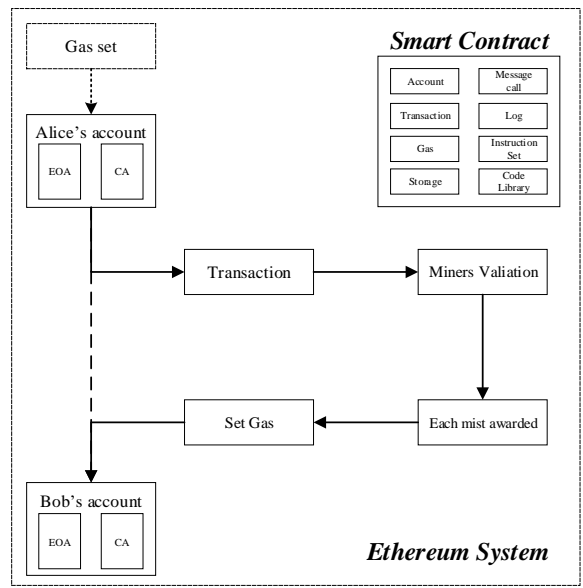

**Fig. 7.** Ethereum Transactions Adding

### 3.3    EOS

EOS (Enterprise Operation System) is a blockchain-based operating system which provides a platform for the development of secure and scalable decentralized applications (DApps) [18]. It provides databases, account permissions, scheduling, authentication and Internet application communication, which greatly improves the efficiency of intelligent business development.

EOS provides not only tools for DApps, but also solutions for scalability issues, which we will further discuss in Section 3.5. EOS has the following three main features [19]:

1. Low latency. The platform supports low latency with DPoS mechanism.
2. Parallel Performance. The off-load can allocated among multiple CPUs and computers in terms of large-scale applications. This avoid heavy on-chain workload.
3. Sequential Performance. Due to some limitations in sequentially dependent steps, those applications that cannot support parallel algorithms will be provided with fast sequential processing for high volumes.

By deploying DPoS (Delegated Proof of Stake) consensus mechanism [20], this permissioned EOS blockchain has become suitable for not only public occasions, but also private enterprise cases. The most representative enterprise cases include:

DPOS is an improved version of PoS for permissioned purposes. DPOS selects nodes (block producers) as representatives to partake in later transaction validation work [21]. At the initial stage of each round, a total of 21 block producers are selected (voted), among which 20 producers are chosen automatically while the remaining one was chosen based on the voting proportion results of other producers. Then these 21 producers will begin to validate blocks of transactions. As long as 15 producers out of 21 reach consensus, a block is considered to be valid.

It is noticeable that the number of selected block producers, 21, is not an absolute unchangeable number. According to latest EOS Whitepaper, the number of super nodes can be voted by the community. However, why is number 21 chosen?

For the comprehensive consideration of efficiency and fairness, the DPoS consensus mechanism set up 21 super nodes as block producers. Firstly, there must be an odd number of nodes, because in EOS whitepaper, there is a "most nodes are just" assumption, as well as a "longest chain mechanism". The odd number of producers can guarantee that only one longest chain exists.

Secondly, the originator, Daniel Larimer, first used 101 witness nodes when making the first version of DPoS consensus mechanism, while in the upgraded version, the number of 101 is changed to user-defined, so that people can freely adjust it when voting. However, when a community is in a controllable state, the number of nodes that can be voted is usually about 15. Therefore, when Daniel conducts the second DPoS project, the number of nodes is set slightly higher than 15 to 21, to ensure the "decentralized" operation under the controllable state. In EOS Whitepaper, there is a confirmation of "absolute irreversibility", which requires the consent of more than 2/3 of the nodes. If the number of nodes is large, a longer waiting time is required for confirmation. If the number of nodes is small, shorter waiting time is prone to some concentration risks. It is understandable that 21 is a balance between decentralization and performance.

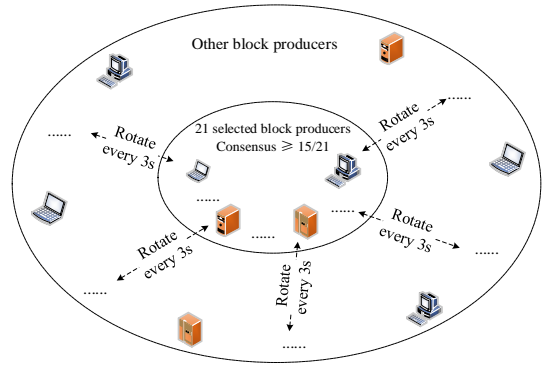

**Fig. 8.** EOS Architecture

There is also rotation mechanism in the selection of producers: every three seconds, the 21 producers are selected from all producers again. This means producers without enough computing power will be sifted out. With no peer competition and shorter consensus confirmation time, it is possible for EOS to improve its scalability and performance of TPS in each unit time [22]. In Fig. 8, the principles of DPoS and the rotation

Especially in a company, this small-scale permissioned stake mechanism allows only some directors, similar to the 21 selected block producers, to have the right in the income, property, copyrights etc. in proportion to investment or token in the account [22]. This can strengthen the administration inside an organization.

### 3.4 Hyperledger Fabric

Hyperledger is a project of open source blockchains to support collaborative development of blockchain- based distributed ledgers. Among them, Hyperledger Fabric is a permissioned blockchain system which aims to build a foundational blockchain platform for enterprises. It provides a modularized framework for enterprises and supports authority management and data security. The two most distinct improvements brought by Hyperledger Fabric are efficiency and confidentiality.

Hyperledger Fabric first introduces the blockchain technology for enterprise use. Compared with blockchain technology, the advantages of Hyperledger Fabric are reflected in the increase in performance and strength on confidentiality. The Hyperledger Fabric architecture is shown in Fig. 9.

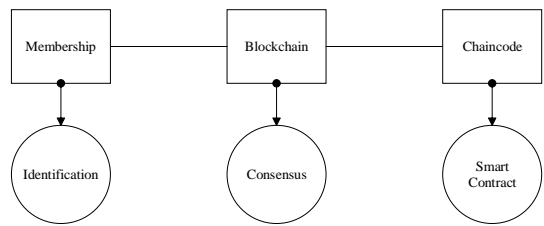

**Fig. 9.** Hyperledger Fabric Architecture

There are three main components in Hyperledger Fabric: Membership, Blockchain and Chaincode. Membership part provides identification services. Blockchain part provides consensus services. Chaincode part is a program that acts as smart contracts in this system. In enterprise scenarios, each node could access this system through the membership services.

The network is permissioned because the participants are known to each other, rather than anonymous and therefore fully untrusted. This is the most distinct difference from the traditional public permissionless Bitcoin and Ethereum blockchain system. The whole system could use general-purpose programming languages such as Java, Go and Node.js, rather than constrained domain-specific languages. However, this uniform programming style and the strict identification process also limit the scalability of the whole system [23].

There are many smart contracts in this system and each maintains a specific type of transaction. Different smart contracts are in charge of different types of transactions. The smart contract will assign endorsers in a specific type of transaction. The endorser is a node which is qualified to validate this specific transaction. The smart contract could also set requirements of completing some specific transaction. For example, it could stipulate that a transaction is completed with validation from 2/3 of endorsers.

When a transaction is initiated, some specific smart contracts will be triggered. Then this transaction will be sent to relevant endorser nodes, which will endorse this transaction. If this transaction is validated, then the result will be directly sent to the user, but not committed on the chain. In this way, the transaction is executed before being validated by the system. Finally, all the transactions, no matter successful or not, will be gathered by the order node for the validation of the whole system. This "execute-order-validate" mechanism is shown in Fig. 10.

This "execute-order-validate" mechanism greatly improves the performance and scalability of the whole system. This first phase also eliminates any non-determinism, as inconsistent results can be filtered out before ordering. Because we have eliminated non-determinism, Fabric is the first blockchain technology that enables use of standard programming languages, which in turn improves the extensibility and scalability of the system. The highest TPS of Fabric could reach 20000 [24].

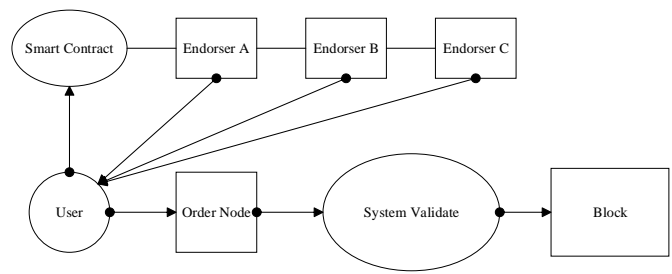

**Fig. 10.** Fabric "execute-order-validate" Mechanism

Another attractive attribute of Fabric is the high confidentiality. The lack of confidentiality can be problematic for many enterprise-use cases, because it is impossible to maintain business relationships in a completely transparent network. Hyperledger Fabric enables confidentiality through its channel architecture and private data feature. The system could set the availability of specific data by assigning authorized peers. The assignment of confidentiality is shown in Fig. 11.

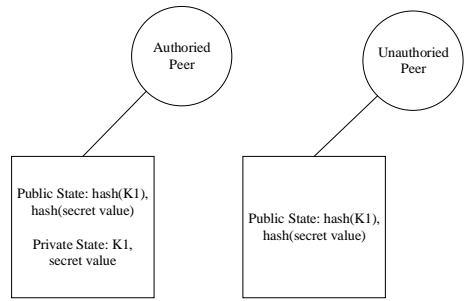

**Fig. 11.** Fabric confidentiality

### 3.5 R3 Corda

R3 Corda [25] was developed to make it easier to record and process financial transactions. It uses a peer-to-peer model in which each peer stores data that relates to all the transactions it has participated in. Consequently, re-creating an audit trail requires querying multiple nodes involved in a chain of transactions. This approach can secure data about transactions by securing the appropriate set of peers. Corda simplifies the creation, automation and enforcement of smart contracts -- a key application of blockchain -- compared to DAG-based distributed ledger technologies. In addition, the Iota Foundation just announced an alpha version of the Iota Smart Contracts Protocol, which could provide functions similar to Corda's.

There are two types of membership in Corda: working node and notary node [26]. The working nodes are in charge of ledger recording as in Blockchain and Ethereum. The notary nodes are trusted by involved parties of transaction and can provide validation of effective transactions. Each notary node is connected with a database or a database cluster. The "effective" here means a certain input data has not been or is not becoming the input of other transactions to ensure that there is no "double spending" issue. Corda is a "permissioned" global network. One working node can be connected to different notary nodes in different transactions, and only involved parties (nodes) will have access and maintain the data of a transaction. Notary nodes will ensure effective transactions and prevent "double spending" issue.

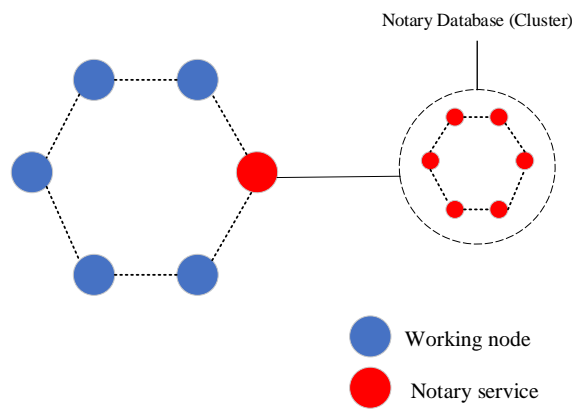

**Fig. 12.** Single Notary Network

There are multiple notary nodes in a notary service to satisfy the requirement of consensus and finally figure out the decision. These notary services are decentralized – each group may have its own notary service as well as consensus algorithm. Shown in Fig. 12 is the basic structure of a Corda network, where more than one working node may be connected to a notary, with each notary consists of more than one notary nodes, consisting of a notary database cluster. This is also called "Single Notary Network".

There are two other kinds of notary service model in Corda: "Clustered Multiple Notary Network" and "Distributed Multiple Notary Network". Different types of notary network are deployed according to the requirements of the financial enterprise system.

DAG, directed acyclic graph [27], is a data structure put forward to improve the TPS of blockchain system. The traditional blockchain consensus mechanism is choosing the longest chain. However, DAG consensus mechanism is choosing the heaviest chain.

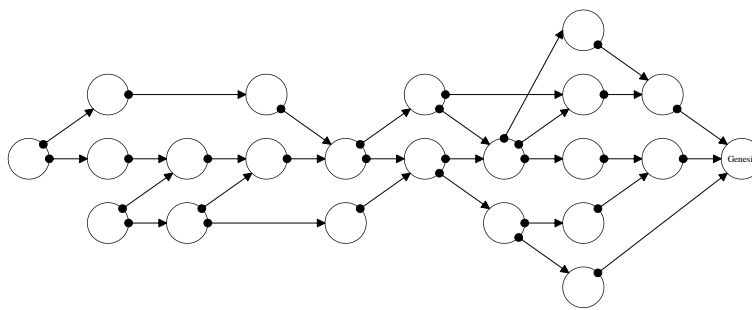

**Fig. 13.** Directed Acyclic Graph

As we can see in Fig. 13, each slot can have more than one legal transaction, and every legal transaction can be verified and added into this system. So, the DAG system can save much time spent on synchronization in traditional blockchain system, because DAG system needn't to synchronize. Considering there can be repeated transactions in each slot, the improvement on TPS is not linear, but implementing this DAG data structure can improve the efficiency of the whole system. The nature of asynchronization also extends the scalability of the system. The difficulty of modification in this system is tremendous because there are many inputs and outputs in one slot and one modification can introduce a series of mistakes, so DAG system can provide users with integrity.

### 3.6 Solana

Solana is a blockchain system which brings tremendous improvement to the performance of traditional blockchain and makes it possible to build scalable and user-friendly applications for the world. It possesses all attributes of traditional blockchain systems but the performance is much better. To improve performance of traditional blockchain system, Solana introduce the Proof of History mechanism.

In Solana system There are two kinds of nodes: Leader and Verifier. The Leader is an elected Proof of History generator, and Solana rotates leaders at fixed intervals. The components of Solana are shown in Fig. 12 The leader will receive the transactions coming from users and order them into a Proof of History sequence.

Proof of History is a mechanism used in Solana. The Proof of History sequence is a list of transactions. The transactions are prearranged by a "Leader", and the timestamp is embodied in this data structure. Every event has a unique hash and account along this data structure. As a function of real time, this information tells us what event had come before another. For example, if we want to know the hash value when index is 300, the only way is to run this algorithm 300 times. We can know that there is real time elapsing in this process from this specific data structure. Time cannot be faked and the future can also not be forecasted. In this way, this system will no longer need to waste computing resources on synchronizing time, because time is preconfigured and unchangeable.

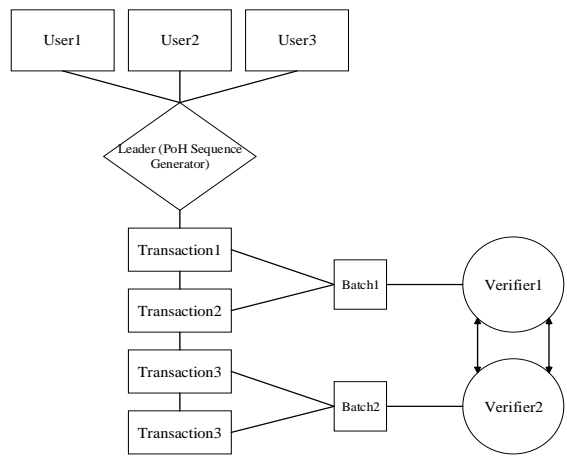

**Fig. 14.** Components of Solana

Then, transactions will be broken into batches. For example, if the leader wants to send 100 transactions to 10 nodes, it would break 100 transactions into 10 batches and send one to each node. This allows the leader to put 100 transactions on the wire, not 100 transactions for each node. Each node then shares its batch with its peers to reconstruct the original collection of 100 transactions. The process of synchronization between verifiers is shown in Fig.13. The combination of Proof of history and horizon scaling can improve the performance tremendously [28].

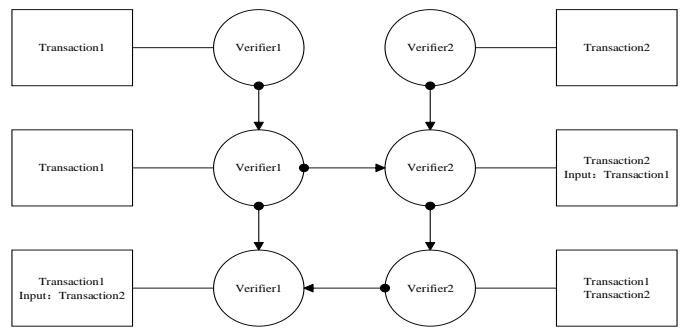

**Fig. 15.** Verifier Synchronization

# 4 Innovating Blockchain Technology for Enterprise Applications

Section 3 presented a quick overview on how blockchain technologies have been evolved, especially over the last 5 years, to satisfy the basic needs for enterprise applications. This section will first help business to select a proper platform that and then proceed to discuss some further research & development directions and options.

## 4.1 Select "Best Proper" to Meet Current Business Needs

Technology has been driving to improve business productivity in almost every industry. But whether certain technology really fits the business has become increasingly critical as it is almost inevitable that technology will be a part of the business or its solution. Any change in the business strategy or model will propagate through the company's technology paradigm especially when the old architecture or organizational structure could not support the new business paradigm.

Every organization need to understand blockchain's unique benefits and focus only on problems that it is best suited to solve. Wrong choices of blockchain platforms can carry significant risks, as they could incur project cost overruns and delays, and miss the opportunity for potential benefits. Table 1 provided some quick guidelines on how to make a "wise" selection.

**Business Model drives the solution.** Blockchain platform will drive the convergence of organizations towards a network-based economy. As companies are more tightly interconnected and rely on business partners to develop, produce, and deliver products and services, they need to integrate resources and capabilities of the involved partners, and engage in joint implementation and utilization of new technologies that are applied and integrated into their business processes.

Blockchain based technology naturally fits in and can inherently connect partners throughout the ecosystem with the required security, enhanced trustworthy and reliability and integrity. Furthermore, blockchain is a very versatile technology and provides the means for customization, as it is not limited to any specific area of application or purpose.

Several most critical issues that enterprises should first consider are:

1. Is permissioned or permissionless blockchain best fit the business model? Most of the successful deployments are on permissioned private blockchains, as organizations really want control over who can participate, and at what capacity.

2. What types of information truly needs the inherent security and integrity mechanisms that the blockchain technology provides? Required operations are computationally very expensive and need participation of many nodes. Therefore, certain combination of data models with different level of security and integrity requirements should be established.

3. To achieve the integrity of the "distributed ledger", what level of consensus is necessary, and who can or should be trusted to provide such expensive operations in the distributed environments. This will help to select or customize the consensus mechanism in order to further improve transaction throughput.

In addition, sample business use cases in Table 1 could be applied as a way to find similar "best-fit" matches.

**Technology Differentiators.** Blockchain technologies will operate across the entire ecosystems, and reveal their benefits ideally on the entire business network. Therefore, the technology fitness, its offered performance and expected impact and characteristics is also key to the success of any business.

As indicated in Table 1, performance and scalability are the dominating factors that limit the applicability of a blockchain platform, especially the public chains. When analyzing business capabilities, quantitative measures such as transaction per second (TPS) for every business transaction, number of concurrent users that the system need to support, and their growth rate, etc. will have to be carefully studied. For most organizations, it may not be possible to develop their own platform or significantly enhance a chosen platform, it is necessary for architects to closely watch the most recent additions of the blockchain platforms and why they are introduced – what specific issues they tried to resolve and of course the results. For example, advances from EOS, CORDA and Solana with new data structures, innovative mechanisms for the required and "sufficient enough" consensus has gradually improved poor TPSs exhibited from the original Bitcoin and Ethereaum by 2-4 orders of magnitude, from single digits to over 60000 TPSs.

## 4.2 Some Key Issues to be Further Investigated and Enhanced

**Data Structures and Algorithms**. From the original block structure to Merkel tree with levels of hashed information, to DAG in Corda, it is evident that significant improvements are still possible by innovating on underlying data structures that take advantage of the representational characteristics of the transactions, especially their identity information or business implications ("smart ID"[29]), business semantics [30], temporal patterns [31], etc. With matching algorithms to enforce security and integrity, they will definitely revolutionize the blockchain technology. In this regard, some self-organizing and potentially self-evolving structures，together with the help from artificial intelligence and machine learning (AI/ML), could be better fit while they could automatically do only what's necessary and sufficient.

With such innovative structures, algorithms can be further researched that take advantage of the full spectrum of analytical, stochastic and optimization, and of course AI/ML methods. Only in this way, the inefficiency of the cumbersome consensus and verification process prevailing in the current blockchain platforms can be eventually solved.

**Data Models and Governance**. As essentially *"every company is a data company",* blockchains potentially generate significant amount of new data to provide the required privacy and security, resilience and irreversibility. If bad data are offered correctly or if the data store contains false information but is offered right, they will all end up on the system.

As some high impact incidences of data loss and breach were reported that could discourage companies from transitioning to blockchains, data governance has become more critical. Poor execution of smart contracts could result in bad automated decision-making that could lead to tremendous business risks. Data privacy still remains as a challenging issue while enterprise blockchain projects need to remedy.

**Performance.** As discussed earlier, the performance of a blockchain could be dominated by the least "powerful" participating node in the network. So as Solana did, how to effectively enforce some minimal standards on certain node capabilities, and further classify nodes into different groups with relevant rights and privileges without sacrificing the integrity assurance, could be appealing. It is even better if we can make such decisions adaptive to the business applications and workloads.

It is also feasible to off-load some heavy processing to a secondary support chain or system, while the main blockchain is only used to record the final result of the transactions. For example, organizations always maintain some lists of "trusted" or "trust-worthy" clients, conducting transactions with those clients do not necessarily need on-time completion of all the complex hashing operations for the entire distributed ledger. Instead results from the off-loaded expensive operations only need to be reflected in the main chain, based on the "trust-worthiness" of the partners. Further-more, such delayed mechanism could be easily designed and implemented with smart contracts!

Potentially, blockchains eliminate the requirement for intermediaries in its stream-lined operations, such as transactions as well as real estate. But this is more of a business problem, as it may introduce changes to the business processes and the interaction patterns that need to be properly addressed from the strategy perspectives.

**Scalability.** To improve scalability, multi-layer or multi-chain systems could be introduced, as discussed before. For example, with the Lightning Network [32] of Bitcoin, a second layer to the main blockchain network is added in order to facilitate faster transactions. Plasma [33] of Ethereum has a parent-child structure, processes the transactions in the child-chain, and records the results in the parent-chain. Sharding [34] groups subsets of nodes into smaller networks or 'shards' that are responsible for the transactions specific to their shard. When offered in conjunction with the proof-of-stake consensus mechanism, such mechanisms have the potential to scale up the application.

As summarized in Table 1, private blockchain offer much better scalability, as the nodes in the network are purposely designed and enabled to process transactions in an environment of trusted parties. Therefore, some hybrid chains effectively combining public chains for certain transactions, while employing private chains for other types of transactions would provide the best combination.

In addition, in almost all known business applications, it is not required to have everyone on the eco-system to participate or contribute to establish and maintain the integrity of the distributed ledger. Therefore, policies or even smart contracts could be utilized to restrict participants.

Technically, workloads can be distributed intelligently to reduce processing needs for more "critical" (either business or technical) nodes.

**Interoperability and Standardization.** Another main challenge is the lack of interoperability among the large number of blockchain networks. Over 6,500 projects adopted a variety of blockchain platforms and solutions with different protocols, programming languages, consensus mechanisms, and privacy measures, while most of those blockchains work in silos and do not communicate with other peer networks. The lack of universal standards and uniformity across blockchain protocols further colluded the situation.

Various projects have initiated to address this problem. Ark uses SmartBridges [35] architecture to bridge the gap of communication between the networks and it claims to offer universal, cross-blockchain transmission and transfer with global interoperability. Cosmos [36] uses the Interblockchain Communication (IBC) protocol [37] to enable blockchain economies to operate outside silos, and transfer files between each other.

The lack of standardization also impacts interoperability and eventually lead to increased costs that make mass adoption difficult. Therefore it is vital to establish industry-wide standards and protocols to help enterprises collaborate on application development, and share blockchain solutions as well as integrate with existing systems.

While the International Organization for Standardisation is currently working on a shared global blockchain standard [38], it will be important that major industry leaders and developer communities proactively participate so that right issues, both business and technical, can be addressed.

**Integration with legacy systems.** Industries were so used to the legacy systems, especially the protocols and processes established in line with their structures. For acceptance and seamless adoption, enterprises are required to integrate them with new blockchain based solutions.

Some solutions started to emerge that enable legacy systems to connect to a blockchain backend. For example, Modex Blockchain Database [39] was designed to help organizations without much exposure in blockchains to relish the potential benefitso and remove the dangers posed by the loss of sensitive data.

**Blockchain as a Service (BAAS).** How can a company integrate the blockchain technology into their business without in-house expertise or experience? BaaS can offer a shortcut by packaging the smart contract technology, blockchains and network infrastructure they run on all "as services". BaaS has emerged as a popular choice because it removes much of the encumbrance of setting up a blockchain.

**Some well-known BaaS players include AWS, IBM, Oracle, VMware and Alibaba.** Amazon Managed Blockchain [40] is a fully managed service that allows enterprises to either join public networks or set up and manage private networks with a competitive blockchain hosting solution. For example, the Hyperledger Fabric solution's existing ordering service can be supported by Amazon QLDB technology, empowering an immutable change log and stronger data storage and security. The IBM Blockchain Platform [41] extends a wide variety of blockchain solutions to customers, from hosting and open-source development assistance to consulting and management services, and it excels in developing and managing solutions for supply chain and manufacturing. Oracle [42] offers a cloud service, an on-premises edition, and a SaaS application for supply chain management, featuring near real-time processing, validation rules and controls in smart contracts, ERP integration, exception tracking, and netting-based settlement. In addition, it is possible to adjust workload and resources to individual business model needs. VMware [43] focuses on ensuring that speed and scalability are possible while also maintaining high levels of security through fault-tolerance preservation and employs a home-grown Scalable Byzantine Fault Tolerance (SBFT), an enterprise-grade consensus engine. Alibaba's Cloud Blockchain as a Service [44] can integrate with its Video DNA service, and makes it possible for users to analyze and trade copyright data for images, video, and audio. It provides innovative end-to-end and chip encryption technologies for security, offers organization, permission, and consortium management capability, chaincode man-agement of smart contracts, and also connection to its CloudMonitor for real-time alerts and monitoring.

Even though all those nice features are marked "as a service", they are still lack of the required standard-based "openness" and "interoperability". Setup, configuration, commitments and conformance to performance, scalability, availability, and sometimes even security and privacy still remains difficult and perplexing.

### 4.3    Alternatives to Bockchain Technology

Despite its promises, blockchain adoption has been very slow. Several alternatives to blockchain that provide better performance have emerged, offering organizations options to reduce costs, simplify development and reduce integration challenges while still able to enjoy some of the core benefits of blockchains.

**Alternative Distributed Ledgers.** A simplified distributed ledger, without the complexities involved with the current blockchain technology, is definitely an alternative for trusted decentralized applications. Several options are available, including Hashgraph, Iota Tangle and R3 Corda.

Iota and Hashgraph use Directed Acyclic Graphs (DAGs) as an alternative data structure for maintaining the ledger，while DAG approach allows an application to write data quickly, and requires permission to conduct certain operations that could slow down the transaction. The applications need to be configured to notify users when conflicts occur, and built-in rules rules to help resolve.

An Iota Tangle stores data across a DAG where each node, or vertex, represents a transaction. The network grows via transactions rather than through a compute-intensive mining process. Iota supports micropayments and transactions across IoT devices. It is mostly decentralized, but it does require a coordinator node that oversees and confirms the addition of new transactions.

Hashgraph also eliminates the need for mining to grow the ledger by utilizing its "gossip about gossip" protocol that network nodes use to share information, come to consensus (another key process in blockchain) and add new transactions to the DAG. As new data is added, an audit trail is also appended to the distributed ledger.

**Centralized ledgers.** Amazon's Quantum Ledger Database simplifies the process of implementing a shared database designed for ledger-like applications that provides a cryptographically verifiable audit trail without all the overhead of a distributed ledger or blockchain. It promises the immutability and verifiability of blockchain combined with the ease and scalability of a traditional cloud service。One thing worth noting is that the blockchain could still be a better option with untrusted players.

**Distributed databases.** Distributed databases offer ome combination of data replication and duplication to ensure data consistency and integrity. For example, the OrbitDB [45] open source project was built on top of a distributed filesystem that allows operation even if one node goes down, and can support the creation of a distributed, peer-to-peer databases, and it enables organizations to develop decentralized applica-tions that run when disconnected from the internet and then sync up with other database nodes when connected. It can also allow data sharing in a way that enforces privacy and provides transparency into how data is being used.

However, for performance and usability reasons, it may still be valuable to keep and manage one highly optimized system of record in a centralized database.

**Decentralized storage.** Decentralized (cloud) storage creates a resilient file storage sharing system by partitioning and encrypting data, distributing it for storage on drives on a peer-to-peer network. IPFS [46] and Storj [47] are such offerings that allow developers to store contents (data, web pages, etc.) with much-reduced bandwidth requirements, improved resilience and less impact of censorship.

Storj is another promising distributed storage technology that allows developers to encrypt files, split them into pieces and then distribute them across a global cloud network. It is directly compatible with Amazon S3 storage tools, which should make it easy for cloud developers to weave into applications without learning new tools.

## 5     Conclusion

It is exciting to live in this wonderful world of technologies while innovations lead to new business opportunities that in turn will present new issues calling for better solutions. This paper quickly surveyed some important issues hindering the broad enterprise adoption for the blockchains, a breakthrough that could be served as the foundation of global business transactions and exchanges, not only eliminating unnecessary intermediaries, but more importantly providing the guaranteed security and integrity of transaction information intrinsically and permanently. After some general description, we analyzed 6 representative blockchain platforms, emphasizing how each evolved to alleviate performance and scalability problems inherent in the original technology structure and algorithm stack. We then presented some quick guidelines on how organizations can select a "best-proper" platform to serve its current and future business needs.

Broad adoption of blockchain still requires significant overhaul in many critical areas, and this paper summarized some of the potential improvement opportunities. As it may take a long time before blockchain technology become mature and stable enough with the necessary transaction throughput, proper scalability and interoperability for enterprise applications, this paper finally presented some alternative technology options.